\documentclass[12pt]{article}
\usepackage[dvips]{graphics}
\usepackage{graphicx}
\usepackage{rotating}                    
\usepackage{subfigure}
\usepackage{setspace}
\newcommand{\x}{\textit{x} }

\newcommand{\p}{\partial}
\newcommand{\be}{\begin{equation}}
\newcommand{\ee}{\end{equation}}
\newcommand{\bea}{\begin{eqnarray}}
\newcommand{\eea}{\end{eqnarray}}
\newcommand{\bes}{\begin{equation*}}
\newcommand{\ees}{\end{equation*}}
\newcommand{\ba}{\begin{array}}
\newcommand{\ea}{\end{array}}

\begin{document}
\title{Solution of polarised singlet DGLAP evolution equations by the method of characteristics}
\author{D. K. Choudhury \thanks{e-mail: dkc@yahoo.co.in }\\Department of Physics, Gauhati University, Guwahati-781014, India \and
  P. K. Sahariah\thanks{e-mail: pksahariah@yahoo.com }\\Department of Physics, Cotton College, Guwahati-781001, India }
\date{}
\maketitle
\begin{abstract}
The polarised singlet coupled DGLAP equations (LO) are transformed by a Taylor series expansion at low $x$ into a pair of partial differential equations in $x$ and $t$($t=lnQ^2/\Lambda^2$). The pair of coupled  partial differential equations is then reduced to canonical form and the resultant system solved by applying the method of characteristics under small $x $ and $t$ approximations. The result is compared with some exact solutions available in the literature.

\end{abstract}

\section{Introduction}

The measurement of the polarized structure functions of the proton by the European Muon Collaboration in 1988 revealed that the structure of the proton is more profound to be understood only from the unpolarized DIS. Their result\cite{EMCPLB206,EMCNPB328}, which is often referred as the 'proton spin problem' indicated that the contribution of the quarks to the nucleon spin is very small. The rest of the spin then must be carried by the gluon and/or by the angular momentum of quarks and gluons. 
The spin distributions among the different contributors is expressed by a sum rule \cite{JaffeManohar,XJi}
\be
\label{eqn:ch6spinsumrule}
\frac{1}{2}=\frac{1}{2} \Delta \Sigma + \Delta G + L_q + L_g \, ,
\ee
where $ \Delta \Sigma $ is the contribution from the quarks, $\Delta G $ from the gluons and $L_q $ , $L_g $ are the orbital angular momentum of quarks and gluons to the total spin of the nucleon. In the static quark model with SU(6) symmetry, the spin is entirely carried by the quarks i.e  $ \Delta \Sigma =1 $. In the naive quark-parton model,  $ \Delta \Sigma $ is predicted to be $\sim 0.7$. These are in contradiction to the  EMC result. Based on the next-to-leading order QCD analysis of the deep inelastic polarized data, there are several model parametrization \cite{AAC00, LSSEPJC23,GRSV01,BlumleinBottcher01,Gehrmanref9,Ghoshetal}  of the polarized parton distribution functions. All of these models predict that the value of  $ \Delta \Sigma $ is around $0.3$ or less; that is, the quark contribution to the nucleon spin is very small contrary to our expectation. In view of this, the polarized gluon distribution has attracted in recent years special interest in polarized DIS, because, a rather large contribution of the first moment of the gluon distribution is essential to compensate for the low value of the quark contribution to the total helicity of the nucleon. Recent data \cite{PLAnthonyPLB493} and analyses \cite{ LSSEPJC23,GRSV01,BlumleinBottcher01,Gehrmanref9,Ghoshetal} suggest that the contribution of the polarized gluon momentum is quite large, though the error involved is too large to significantly constrain its contribution. It is therefore imperative to know how the gluon is polarized inside the nucleon.
In this paper we investigate the polarized gluon and the quark singlet contribution to the nucleon spin with the help of the DGLAP evolution equations. We adopt the method of characteristics  to solve the coupled polarized DGLAP evolution equations. In  \S \ref{sec:formalism} we discuss the formalism and in \S \ref{sec:results} we discuss our results.

\section{Formalism}
\label{sec:formalism}
For $Q^2$ evolution it is convenient to introduce a flavour singlet combination of polarized quark and anti-quarks   by
\be
\label{eqn:ch6delsigma}
\Delta \Sigma (x,Q^2) = \sum_q ( \Delta q(x,Q^2)+\Delta \bar{ q}(x,Q^2)),
\ee
where the sum is over all quark and anti-quark flavours. Here $\Delta q(x,Q^2)=(q^+(x,Q^2)-q^-(x,Q^2)), $ where $ q^+(x,Q^2)$ ,( $q^-(x,Q^2) $) refers to the quark distribution with helicity parallel (anti-parallel) to the parent proton. The polarized gluon distribution is similarly defined as 
\be
\label{eqn:ch6polgluon}
\Delta g(x,Q^2)=g^{(+)}(x,Q^2)-g^{(-)}(x,Q^2).
\ee
The $Q^2$ evolution of the polarized singlet and the gluon distribution  mix non-trivially and is given by the DGLAP \cite{Gribov,Dokshitzer,AP} equations 
\be
\label{eqn:ch6singcoup}
\frac{\p}{\p t}\left(\ba{c}
\Delta \Sigma\left(x,t\right)\\
\Delta g\left(x,t\right)
\ea
\right) = \frac{\alpha_s\left(t\right)}{2\pi}\left(
\ba{cc}
\Delta P_{qq} & 2\,n_f\,\Delta P_{qg} \\
\Delta P_{gq} & \Delta P_{gg}
\ea
\right)\otimes \left(
\ba{c}
\Delta \Sigma\left(x,t\right) \\
\Delta g\left(x,t\right)
\ea
\right),
\ee
where $\,t=ln \frac{Q^2}{\Lambda^2}\,$ and $ \Delta P_{ij} $ are polarized splitting functions which are known at LO \cite{AP} and NLO \cite{NLOPolspfn1,NLOPolspfn2,NLOPolspfn3}. The symbol $\otimes$ stands for the usual Mellin convolution in the first variable defined as
\be
\label{eqn:ch6convolution}
(\Delta P \times \Delta q)(x,t)=\int_x^1\,\frac{dy}{y}\Delta P\left(\frac{x}{y}\right)\, \Delta q(y,t).
\ee
 The polarized parton distributions at an initial scale enter as boundary values in the solution of the above equations. However, with exact splitting functions,  analytic solutions in the entire range of $ x $ is not possible. The splitting functions can be simplified by taking their moments and expanding around the rightmost singularity at $N=0 $ where $N$ is the moment variable. In this procedure  the splitting functions\cite{Gehrmanref9} at LO get simplified  to
\begin{eqnarray}
\label{eqn:ch6splittingfn}
\left. \begin{array}{cc}
\hspace{-0.2in}\Delta P_{qq}^{(0)}(x)=\frac{4}{3}[1+\frac{1}{2}\delta (1-x)] ,\hspace{-0.25in} & \Delta P_{qg}^{(0)}(x)=\frac{1}{2}\,2 n_f[-1+2\delta (1-x)], \\ 
\hspace{-0.15in}\Delta P_{gq}^{(0)}(x)=\frac{4}{3}[2-\delta (1-x)] ,\hspace{-0.1in} & \Delta P_{gg}^{(0)}(x)=3[4-\frac{13}{6}\delta (1-x)]-\frac{n_f}{3}\delta (1-x)
\end{array}\right\}.
\end{eqnarray}
 With these simplified splitting functions, Eqs.(\ref{eqn:ch6singcoup}) can be solved analytically with some  approximations about small $x$ behaviour of these to be discussed later.
The two equations  Eqs.(\ref{eqn:ch6singcoup}) can be written separately in LO as
\be
\label{eqn:ch6polquark}
\frac{\p}{\p t}\Delta \Sigma (x,t)=\frac{\alpha_s(t)}{2 \pi}\left[\Delta P_{qq}^{(0)}\otimes \Delta \Sigma (x,t)+2\,n_f \Delta P_{qg}^{(0)}(x) \otimes \Delta g(x,t)\right]
\ee
and
\be
\label{eqn:ch6polgluonxt}
\frac{\p}{\p t}\Delta g (x,t)=\frac{\alpha_s(t)}{2 \pi}\left[\Delta P_{gq}^{(0)}\otimes \Delta \Sigma (x,t)+ \Delta P_{gg}^{(0)}(x) \otimes \Delta g(x,t)\right]\,.
\ee
Using the convolution integral(Eq.\ref{eqn:ch6convolution}), we write Eq.(\ref{eqn:ch6polquark}) and Eq.(\ref{eqn:ch6polgluonxt}) as
 \be
\label{eqn:ch6polquark1}
\frac{\p}{\p t}\Delta \Sigma (x,t)=\frac{\alpha_s(t)}{2 \pi}\left[\int_x^1dz\, \Delta P_{qq}^{(0)}(z)\, \Delta \Sigma (\frac{x}{z},t)+2\,n_f \int_x^1dz\, \Delta P_{qg}^{(0)}(z) \, \Delta g(\frac{x}{z},t)\right],
\ee
\be
\label{eqn:ch6polgluon1}
\frac{\p}{\p t}\Delta g (x,t)=\frac{\alpha_s(t)}{2 \pi}\left[\int_x^1dz\, \Delta P_{gq}^{(0)}(z)\, \Delta \Sigma (\frac{x}{z},t)+ \int_x^1dz\, \Delta P_{gg}^{(0)}(z)\,  \Delta g(\frac{x}{z},t)\right].
\ee
Introducing a variable $u$ as $u=1-z$, we can write $\displaystyle{\frac{x}{z}}$ as a series
\be
\label{eqn:xbyzseries}
\frac{x}{z}=\frac{x}{1-u}=x\sum_{k=0}^{\infty }u^{k} .
\ee
Now we expand $\displaystyle{\Delta \Sigma (\frac{x}{z},t)} $ and $\displaystyle{\Delta g(\frac{x}{z},t)} $ in Eq.(\ref{eqn:ch6polquark1}) and Eq.(\ref{eqn:ch6polgluon1}) in Taylor series using Eq.(\ref{eqn:xbyzseries}) as
\begin{equation}
\label{eqn:ch6Taylorsigma}
\Delta \Sigma (\frac{x}{z},t)=\Delta \Sigma (x+x\sum_{k=1}^{\infty }u^{k},t) 
=\Delta \Sigma(x,t)+(x\sum_{k=1}^{\infty }u^{k})\frac{\partial \Delta \Sigma(x,t)}{%
\partial x}+ ...
\end{equation}
and 
\begin{equation}
\label{eqn:ch6Taylorgluon}
\Delta g\, (\frac{x}{z},t)=\Delta g (x+x\sum_{k=1}^{\infty }u^{k},t) 
=\Delta g(x,t)+(x\sum_{k=1}^{\infty }u^{k})\frac{\partial \Delta g(x,t)}{%
\partial x}+....\,  .
\end{equation}
At low $x\,  $ the above series Eq.(\ref{eqn:ch6Taylorsigma}) and Eq.(\ref{eqn:ch6Taylorgluon}) are convergent\cite{JSDKCGMPLB403}. Hence, neglecting higher order terms, the series can be approximated as
\begin{equation}
\label{eqn:ch6Taylorsigmaapprox}
\Delta \Sigma (\frac{x}{z},t)
\approx \Delta \Sigma(x,t)+(x\sum_{k=1}^{\infty }u^{k})\frac{\partial \Delta \Sigma(x,t)}{%
\partial x}
\end{equation}
and 
\begin{equation}
\label{eqn:ch6Taylorgluonapprox}
\Delta g (\frac{x}{z},t) 
\approx \Delta g(x,t)+(x\sum_{k=1}^{\infty }u^{k})\frac{\partial \Delta g(x,t)}{%
\partial x}.
\end{equation}
Using  Eq.(\ref{eqn:ch6Taylorsigmaapprox}) and  Eq.(\ref{eqn:ch6Taylorgluonapprox}) and the splitting functions given by Eqs.(\ref{eqn:ch6splittingfn}), we can write Eq.(\ref{eqn:ch6polquark1}) and Eq.(\ref{eqn:ch6polgluon1})  as
\be
\label{eqn:ch6polquark2}
\frac{\beta_0 t}{2}\frac{\p \Delta \Sigma (x,t)}{\p t}=I_q^{(1)}(x) \Delta \Sigma (x,t)+I_q^{(2)}\frac{\p \Delta \Sigma (x,t)}{\p x}+I_g^{(1)}(x) \Delta g (x,t)+I_g^{(2)}\,\frac{\p \Delta g(x,t)}{\p x}
\ee
and
\be
\label{eqn:ch6polgluon2}
\frac{\beta_0 t}{2}\frac{\p \Delta g (x,t)}{\p t}=I_q^{(3)}(x) \Delta \Sigma (x,t)+I_q^{(4)}\frac{\p \Delta \Sigma (x,t)}{\p x}+I_g^{(3)}(x) \Delta g (x,t)+I_g^{(4)}\,\frac{\p \Delta g(x,t)}{\p x}
\ee
where we have used $\displaystyle{ \alpha_s(t)=\frac{4\,\pi}{\beta_0 \, t}}$ in LO. The quantities $I_i^j (i=q,g;j=1,2,3,4 )$ in Eq.(\ref{eqn:ch6polquark2}) and Eq.(\ref{eqn:ch6polgluon2}) are given by
\be
\label{eqn:ch6Iq1}
\hspace{-0.9in}I_q^{(1)}(x)=\frac{4}{3}\int_x^1\frac{dz}{z}\left(1+\frac{1}{2}\delta(1-z)\right)\, ,  
\ee
\be
\label{eqn:ch6Iq2}
\hspace{-0.3in}I_q^{(2)}(x)=\frac{4}{3}\int_x^1\frac{dz}{z}\left(1+\frac{1}{2}\delta(1-z)\right)\left(x\,\sum_{k=1}^{\infty}u^k\right)\, ,
\ee
\be
\label{eqn:ch6Ig1}
\hspace{-0.7in}I_g^{(1)}(x)=2\,n_f\frac{1}{2}\int_x^1\frac{dz}{z}\left(-1+2\,\delta(1-z)\right)\, ,
\ee
\be
\label{eqn:ch6Ig2}
I_g^{(2)}(x)=2\,n_f\frac{1}{2}\int_x^1\frac{dz}{z}\left(-1+2\,\delta(1-z)\right)\left(x\,\sum_{k=1}^{\infty}u^k\right)\, ,
\ee
\be
\label{eqn:ch6Iq3}
\hspace{-1.2in}I_q^{(3)}(x)=\frac{4}{3}\int_x^1\frac{dz}{z}\left(2-\delta(1-z)\right)\, ,
\ee
\be
\label{eqn:ch6Iq4}
\hspace{-0.5in}I_q^{(4)}(x)=\frac{4}{3}\int_x^1\frac{dz}{z}\left(2-\delta(1-z)\right)\left(x\,\sum_{k=1}^{\infty}u^k\right)\, ,
\ee
\be
\label{eqn:ch6Ig3}
\hspace{-0.10in}I_g^{(3)}(x)=\int_x^1\,\frac{dz}{z}\left[3\left(4-\frac{13}{6}\delta(1-z)\right)-\frac{n_f}{3}\delta(1-z)\right]
\ee
and
\be
\label{eqn:ch6Ig4}
\hspace{-0.3in}I_g^{(4)}(x)=\int_x^1\,\frac{dz}{z}\left[3\left(4-\frac{13}{6}\delta(1-z)\right)-\frac{n_f}{3}\delta(1-z)\right]\,\left(x\,\sum_{k=1}^{\infty}u^k\right)\, .
\ee
Carrying out the integrations in Eqs.(\ref{eqn:ch6Iq1}-\ref{eqn:ch6Ig4}), we recast  Eq.(\ref{eqn:ch6polquark2}) and Eq.(\ref{eqn:ch6polgluon2}) as two coupled partial differential equations in two variables $x$ and $t$ as :
\begin{eqnarray}
\label{eqn:ch6pde1}
\hspace{-0.2in} a_{11}^\prime\frac{\p \Delta \Sigma(x,t)}{\p t}+a_{12}^\prime\frac{\p \Delta g(x,t)}{\p t}+b_{11}^\prime\frac{\p \Delta \Sigma(x,t)}{\p x}+b_{12}^\prime\frac{\p \Delta g(x,t)}{\p x} \nonumber \\
=R_{11}^\prime\Delta \Sigma(x,t)+R_{12}^\prime\Delta g(x,t) 
\end{eqnarray}
and
\begin{eqnarray}
\label{eqn:ch6pde2}
\hspace{-0.2in} a_{21}^\prime\frac{\p \Delta \Sigma(x,t)}{\p t}+a_{22}^\prime\frac{\p \Delta g(x,t)}{\p t}+b_{21}^\prime\frac{\p \Delta \Sigma(x,t)}{\p x}+b_{22}^\prime\frac{\p \Delta g(x,t)}{\p x} \nonumber \\
=R_{21}^\prime\Delta \Sigma(x,t)+R_{22}^\prime\Delta g(x,t), 
\end{eqnarray}
where
\begin{eqnarray}
\label{eqn:ch6aelements}
\left. \begin{array}{cc}
a_{11}^\prime=t , & a_{12}^\prime=0 \\
a_{21}^\prime=0 , & a_{22}^\prime=t
\end{array}\right\}\, ,
\end{eqnarray}
\begin{eqnarray}
\label{eqn:ch6belements}
\left. \begin{array}{cc}
b_{11}^\prime=-\frac{4}{3}\left(1-x-x\,ln\frac{1}{x}\right) , & \hspace{0.3in}  b_{12}^\prime=-2\,n_f\,\frac{1}{2}\left(-1+x+x\,ln\frac{1}{x}\right) \\
b_{21}^\prime=-\frac{4}{3}\left(1-2 x-x\,ln\frac{1}{x}\right) , &  b_{22}^\prime=-12\left(1-x-x\,ln\frac{1}{x}\right)
\end{array}\right\}
\end{eqnarray}
and
\begin{eqnarray}
\label{eqn:ch6Relements}
\left. \begin{array}{cc}
\hspace{-0.3in}R_{11}^\prime=\frac{4}{3}\left(\frac{1}{2}+ln\frac{1}{x}\right), \hspace{1.0in} R_{12}^\prime=2\,n_f\,\frac{1}{2}\left(2-ln\frac{1}{x}\right) \\
R_{21}^\prime=\frac{4}{3}\left(2\,ln\frac{1}{x}-1\right),\hspace{1.0in}  R_{22}^\prime=\left[12\,ln\frac{1}{x}-\left(\frac{13}{2}+\frac{n_f}{3}\right)\right]
\end{array} \right\}\, .
\end{eqnarray}
Eq.(\ref{eqn:ch6pde1}) and Eq.(\ref{eqn:ch6pde2}) represent a system of first order coupled partial differential equations. We can reduce these equations to canonical form and then solve by the method of characteristics. To do that, we introduce a vector $\vec{\Delta u}(x,t) $ defined  by
\be
\label{eqn:ch6vecdelu}
\vec{\Delta u}(x,t)=\left(\ba{c}
\Delta \Sigma (x,t)  \\
\Delta g(x,t)
\ea
\right)
\ee
and express the two equations Eq.(\ref{eqn:ch6pde1}) and Eq.(\ref{eqn:ch6pde2}) in matrix form
\be
\label{eqn:ch6pdemf}
a' \vec{\Delta u}_t(x,t)+b'\vec{\Delta u}_x(x,t)=R'\vec{\Delta u}(x,t),
\ee
where the matrices $a' $\,, $ b' $ and $ R' $ are 
\be
\label{eqn:ch6mata}
a'=\left(\ba{cc}
a_{11}^\prime & a_{12}^\prime \\
a_{21}^\prime & a_{22}^\prime
\ea
\right)\, ,
\ee
\be
\label{eqn:ch6matb}
b'=\left(\ba{cc}
b_{11}^\prime & b_{12}^\prime \\
b_{21}^\prime & b_{22}^\prime
\ea
\right)\hspace{0.13in}
\ee
and
\be
\label{eqn:ch6matR}
R'=\left(\ba{cc}
R_{11}^\prime & R_{12}^\prime \\
R_{21}^\prime& R_{22}^\prime
\ea
\right),
\ee
the elements being given by Eqs.(\ref{eqn:ch6aelements}-\ref{eqn:ch6Relements}) respectively. The matrix $ a' $ being non-singular, we multiply Eq.(\ref{eqn:ch6pdemf}) from left by $ a'^{-1} $  and get
\be
\label{eqn:ch6pdeAB}
 \vec{ \Delta u}_t(x,t)+A'\,\vec{\Delta u}_x(x,t)=B'\,\vec{\Delta u}(x,t)\, ,
\ee
where the new matrices $A'$ and $B'$ are:
\be
\label{eqn:ch6A}
A'={a'}^{-1}b'
\ee
and
\be
\label{eqn:ch6B}
B'={a'}^{-1}R' .
\ee
Eq.(\ref{eqn:ch6pdemf}) represents  a system of two coupled first order partial differential equation in the two variables \x and \textit{t}  for the vector $  \vec{\Delta u} $ prescribed by Eq.(\ref{eqn:ch6vecdelu}). Its principal part, i.e. $ \vec{\Delta u}_t(x,t)+A'\vec{\Delta u}_x(x,t) $ is completely characterized by the coefficient matrix $A'$. Since the matrix $ A'$ has n [here n=2] distinct eigenvalues, the system Eq.(\ref{eqn:ch6pdeAB} ) is a hyperbolic one and it is possible to obtain its canonical form in the following way \cite{Zachmanoglou,Farlow}: Let ${\lambda}'^{(i)} (i=1,2) $ be the eigenvalues of the matrix $A' $ (Eq.\ref{eqn:ch6A}) and  $ P'$ be the matrix formed by the corresponding eigenvectors.
Now, if $ \Lambda $ is the diagonal matrix with the eigenvalues $\lambda_1$ and $ \lambda_2$ as the two elements, then we have
\be
\label{eqn:ch6PAP}
P'^{-1}A'P'\equiv\Lambda=\left(\ba{cc}
\lambda_1 & 0 \\
0 & \lambda'_2 
\ea \right).
\ee
 Let $\vec{\Delta v}(x,t)$ be a new function defined in terms of $\vec{\Delta u}(x,t)$ by the equation
\be
\label{eqn:ch6vdef}
\Delta \vec{u}(x,t)=P'.\Delta \vec{v}(x,t)\, .
\ee
so that 
\be
\label{eqn:ch6upv}
\vec{\Delta v}={P'}^{-1}.\vec{\Delta u}.
\ee
Differentiating Eq.(\ref{eqn:ch6vdef}) with respect to $ t$ and $x$ respectively we get
\be
\label{eqn:ch6utux}
\vec{\Delta u}_t=P'\vec{\Delta v}_t+P'_t \vec{\Delta v} , \hspace{0.4in}\vec{\Delta u}_x=P'\vec{\Delta v}_x+P'_x \vec{\Delta v}.
\ee
Substituting Eqs.(\ref{eqn:ch6vdef}) and (\ref{eqn:ch6utux}) in Eq.(\ref{eqn:ch6pdemf}), we obtain
\be
\label{eqn:ch6ptpx}
P'\vec{\Delta v}_t+P'_t \vec{\Delta v}+A' P'\vec{\Delta v}_x+A' P'_x \vec{\Delta v}=B'P'\vec{\Delta v}.
\ee
Multiplying Eq.(\ref{eqn:ch6ptpx}) from left by $ P'^{-1}$ and using Eq.(\ref{eqn:ch6PAP}), we obtain 
\be
\label{eqn:ch6vtvx}
\vec{\Delta v}_t+\Lambda \vec{\Delta v}_x=\vec{\Delta e},
\ee
where
\be
\label{eqn:ch6vece}
\vec{\Delta e}=P'^{-1}(B'.P'-P'_t-A'.P'_x). \vec{\Delta v}\, .
\ee
Eq.(\ref{eqn:ch6vtvx}) is in canonical form. In components, it is 

\be
\label{eqn:ch6canonicalform}
\Delta v_t^{(i)}+{\lambda}'^{(i)} \,\Delta v_x^{(i)}=\Delta e^{(i)} \hspace{0.7in} (i=1,2),
\ee
Eq.(\ref{eqn:ch6canonicalform}) shows that the principal part  viz. $\displaystyle{\frac{\p \Delta v^{(i)}}{\p t}+\lambda'^{(i)} \frac{\p \Delta v^{(i)}}{\p x}}$ involves only the component $\Delta v^{(i)}$ of the vector $\vec{\Delta v}$ and its derivatives i.e. the equations are decoupled.
These equations (Eq.\ref{eqn:ch6canonicalform}) can be reduced to ordinary differential equations
\be
\label{eqn:ch6ode}
\frac{d \Delta v^{(i)}}{d t}=\Delta e^{(i)}(x,t,\Delta v^{(i)}(x,t)) \hspace{0.3in}  (i=1,2) \hspace{-0.4in}
\ee
 along the characteristic curves defined by the equations
\be
\label{eqn:ch6cheqn}
{\lambda}'^{(i)}(x,t)=\frac{d x^{(i)}(t)}{d t}\, .
\ee
In order to integrate the ordinary differential equations (Eq.\ref{eqn:ch6ode}) to get analytical forms for $\vec{\Delta v}$, one has to express the right hand side in terms of $t$. This is done from the solution of the characteristic equations Eq.(\ref{eqn:ch6cheqn}), which expresses $x^{(i)}$ as a function of $t$. However, the integration of Eq.(\ref{eqn:ch6cheqn}) to get such an expression depends on the nature of the eigenvalues ${\lambda}'^{(i)}$ of the matrix $ A' $.  For simple forms of the eigenvalues, it is possible to integrate Eq.(\ref{eqn:ch6ode}) analytically. We discuss this below.

The eigenvalues of the matrix $ A' $ are 
\be
\label{eqn:ch6eienval}
{\lambda}'^{(1,2)}=\frac{1}{9\,t}(b_{11}^\prime+b_{22}^\prime\pm s),
\ee
where
\be
\label{eqn:ch6s}
s=\sqrt{{b'}^2_{11}+4\, b_{12}^\prime\, b_{21}^\prime-2 b_{11}^\prime\,b_{22}^\prime+{b'}^2_{22}}\, .
\ee
Taking the elements of the matrices $ a' $  and $ b' $  as given by Eqs.(\ref{eqn:ch6aelements}-\ref{eqn:ch6belements}), the eigenvalues are found to be 
\begin{eqnarray}
\label{eqn:ch6eigenvalue1}
{\lambda}'^{(1,2)}(x,t)=&&\frac{2}{27\,t}\left[\left(-10+10\,x+10\,x\,ln\frac{1}{x}\right)\right.\nonumber \\
&&\left.\pm \sqrt{55-101\,x+46\,x^2+(-110\,x+101\,x^2)\,ln\frac{1}{x}+55\,x^2\,(ln\frac{1}{x}^2})^2\right] ,\nonumber \\
\end{eqnarray}
where the $ +$ ($-$) sign corresponds to ${\lambda}'^{(1)}$(${\lambda}'^{(2)}$).
And the two components of the vector $\vec{\Delta e}$ are
\bea
\label{eqn:ch6componente1e2}
&\hspace{-4.5in} \Delta e^{(1,2)} =  \nonumber \\ 
& \frac{1}{t}\left[\left(\mp\frac{(-b_{11}^\prime+b_{22}^\prime+s)\left(-\frac{b_{21}^\prime R_{11}^\prime}{ s}-\frac{(-b_{11}^\prime+b_{22}^\prime \mp s)R_{21}^\prime}{2 s }\right)}{9 b_{21}^\prime} \mp \frac{2 b_{21}^\prime R_{12}^\prime}{9 s }\mp \frac{(-b_{11}^\prime+b_{22}^\prime \mp s)R_{22}^\prime}{9 s }\right)\Delta v_1 \right. \nonumber   \\
& \left. +\left(\mp \frac{(-b_{11}^\prime+b_{22}^\prime-s)\left(-\frac{b_{21}^\prime R_{11}^\prime}{s}-\frac{(-b_{11}^\prime+b_{22}^\prime \mp s)R_{21}^\prime}{2 s }\right)}{9 b_{21}^\prime} \mp \frac{2 b_{21}^\prime R_{12}^\prime}{9 s }\mp \frac{(-b_{11}^\prime+b_{22}^\prime \mp s)R_{22}^\prime}{9 s }\right)\Delta v_2 \right]\, ,\hspace{0.2in}
\eea
where the upper sign (-) corresponds to $\Delta e^{(1)}$ and the lower sign (+) to  $\Delta e^{(2)}$.
We note that both ${\lambda}'^{(i)} $ and $\Delta e^{(i)}$  are factorisable in $x$ and $t$, the quantities $b'_{ij} $ and $R'_{ij}$ being functions of $x$ only. As we have already observed, integration of Eq.(\ref{eqn:ch6ode}) depends on the analytical solution of the characteristic equation . But with the above eigenvalues (Eq.(\ref{eqn:ch6eigenvalue1})), analytical solution of the characteristic equations(Eq.(\ref{eqn:ch6cheqn})) cannot be found and so we cannot find the necessary transformation equation to express $x$ as a function of $t$ for the integration of  Eq.(\ref{eqn:ch6ode}). But under certain  extreme situations, it is possible to get  analytical solutions which we discuss below.
\subsection { Solution when $ x \rightarrow 0 $}
The eigen values of the matrix $ A' $ can be simplified by making certain approximations about the elements of the matrix $b'$. Before we do that, we write the eigenvalues in simple form. The eigenvalues being factorisable in $ x $ and $t$ can be written as
\be
\label{eqn:ch6eivalfactorized}
{\lambda}'^{(i)}(x,t)={\lambda}'^{(ia)}(x)\,{\lambda}'^{(ib)}(t)\hspace{0.5in} (i=1,2),
\ee 
where
\begin{eqnarray}
\label{eqn:ch6eivalfactorized1}
{\lambda}'^{(ia)}(x)=&&\frac{2}{27}\left[\left(-10+10\,x+10\,x\,ln\frac{1}{x}\right)\right. \nonumber \\
&&\left.\pm \sqrt{55-101\,x+46\,x^2+(-110\,x+101\,x^2)\,ln\frac{1}{x}+55\,x^2\,ln\frac{1}{x}^2}\right]\hspace{0.3in}
\end{eqnarray}
and
\be
\label{eqn:ch6eivalfactorized2}
{\lambda}'^{(ib)}(t)=\frac{1}{t}\,.
\ee
 In Eq.(\ref{eqn:ch6eivalfactorized1}), $ + $(-) sign  corresponds to i=1 (2) respectively.
Now in the limit   $x \rightarrow  0$,  Eq.(\ref{eqn:ch6eivalfactorized1}) takes the values
\be
\label{eqn:ch6eivallimit}
{\lambda}'^{(ia)} \rightarrow \frac{2}{27}\,(-10\pm \sqrt{55})\,.
\ee
The characteristic equation Eq.(\ref{eqn:ch6cheqn}) now takes the form
\be
\label{eqn:ch6cheqnfactorized}
\frac{d x^{(i)}(t)}{{\lambda}'^{(ia)}(x)}={\lambda}'^{(ib)}(t)dt\, .
\ee
Integrating  Eq.(\ref{eqn:ch6cheqnfactorized})  we get
\be
\label{eqn:ch6cheqnapp}
x^{(i)}(t)=\alpha^{(i)}\,ln t +C^i \, ,
\ee
where
\be
\label{eqn:ch6alpha1}
\alpha^{(i)}=\frac{2}{27}(-10\pm\sqrt{55})\, ,\hspace{1.0in} i=1,2
\ee
and $C^{(i)} $  are two constants of integration. To get the constants of integration, let ($\bar{x},\bar{t}$) be a fixed point \cite{Zachmanoglou} in the ($x,t$) plane through which the two characteristic curves Eq.(\ref{eqn:ch6cheqnapp}) pass through. That is, $x^{(1)}(\bar{t})=\bar{x}$ and  $x^{(2)}(\bar{t})=\bar{x}$. Then from the Eqs.(\ref{eqn:ch6cheqnapp}) we get
\be
\label{eqn:ch6cheqn12}
x^{(i)}(t)=\bar{x}+\alpha^{(i)}\, ln\left(\frac{t}{\bar{t}}\right)\, .\hspace{1.0in} (i=1,2)
\ee
These are the two characteristic curves that pass through a common point ($ \bar{x},\bar{t}$) up to which we can evolve the two functions $\Delta v^{(1)} $ and $ \Delta v^{(2)} $ defined by Eq.(\ref{eqn:ch6upv}). Furthermore, if the two characteristic curves cut the initial line $t=t_0$ $( t_0=ln\frac{Q_0^2}{\Lambda^2})$ at $x^{(1)}(t_0)={\tau}'_1 $ and $ x^{(2)}(t_0)={\tau}'_2 $ respectively, then
\be
{\tau}'_1=\bar{x}+\alpha^{(1)}\,ln \left(\frac{t_0}{\bar{t}}\right)
\ee
and
\be
{\tau}'_2=\bar{x}+\alpha^{(2)}\,ln \left(\frac{t_0}{\bar{t}}\right)\, .
\ee
Hence the two characteristic equations corresponding to the two eigenvalues are respectively
\be
x^{(1)}(t)={\tau}'_1+\alpha^{(1)}\,ln \left(\frac{t}{t_0}\right)
\ee
and
\be
x^{(2)}(t)={\tau}'_2+\alpha^{(2)}\,ln \left(\frac{t}{t_0}\right)\,.
\ee
In this approximation, the two components of the vector $\vec{\Delta e}$ defined in Eq.(\ref{eqn:ch6vece}) are
\bea
\label{eqn:ch6e1}
\Delta e^{(1)}=\frac{1}{2970}\left[(-2255+ 284\sqrt{55})\Delta v^{(1)}+3(275- 52\sqrt{55})\Delta v^{(2)} \right. \\ \nonumber
\left.+(4400-404\sqrt{55})\Delta v^{(1)}+4(-880+119\sqrt{55})\Delta v^{(2)}ln\frac{1}{x^{(1)}(t)}\right]\frac{1}{t}
\eea
and
\bea
\label{eqn:ch6e2}
\Delta e^{(2)}=\frac{1}{2970}\left[-(2255+ 284\sqrt{55})\Delta v^{(2)}+3(275+52\sqrt{55})\Delta v^{(1)}\right. \\ \nonumber
\left.+{(4400+404\sqrt{55})\Delta v^{(2)}-4(880+119\sqrt{55})\Delta v^{(1)}}ln\frac{1}{x^{(2)}(t)}\right]\frac{1}{t}\,.
\eea
Using Eq.(\ref{eqn:ch6e1}) and Eq.(\ref{eqn:ch6e2}), we  rewrite the two differential equations Eqs.(\ref{eqn:ch6ode}) separately as
\be
\label{eqn:ch6odev1}
\frac{d \Delta  v^{(1)}(x,t)}{\Delta v^{(1)}(x,t)}=[A_1(x,t)+B_1(x,t)ln\frac{1}{x^{(1)}(t)}]\frac{1}{t} dt
\ee
and
\be
\label{eqn:ch6odev2}
\frac{d \Delta v^{(2)}(x,t)}{\Delta v^{(2)}(x,t)}=[A_2(x,t)+B_2(x,t)ln\frac{1}{x^{(2)}(t)}]\frac{1}{t} dt \, ,
\ee
where
\be
\label{eqn:ch6A1}
A_1(x,t)=\frac{1}{2970}\left[(-2255+ 284\sqrt{55})+3(275- 52\sqrt{55})\frac{\Delta v^{(2)}}{\Delta v^{(1)}}\right]\,,
\ee
\be
\label{eqn:ch6B1}
B_1(x,t)=\frac{1}{2970}\left[(4400-404\sqrt{55})+4(-880+119\sqrt{55})\frac{\Delta v^{(2)}}{\Delta v^{(1)}}\right]\,,
\ee
\be
\label{eqn:ch6A2}
A_2(x,t)=\frac{1}{2970}\left[-(2255+ 284\sqrt{55})+3(275+52\sqrt{55})\frac{\Delta v^{(1)}}{\Delta v^{(2)}}\right]
\ee
and
\be
\label{eqn:ch6B2}
B_2(x,t)=\frac{1}{2970}\left[(4400+ 404\sqrt{55})-4(880+119\sqrt{55})\frac{\Delta v^{(1)}}{\Delta v^{(2)}}\right]\, .
\ee
 Eq.(\ref{eqn:ch6odev1}) and Eq.(\ref{eqn:ch6odev2}) can now be integrated. Integrating along the characteristics from $t=t_0$ to $t=\bar{t}$ we get
\be
\label{eqn:ch6delv1tbar}
\Delta v^{(1)}(\bar{x},\bar{t})=\Delta v^{(1)}({\tau}'_1)\,\exp\left[\int_{t_0}^{\bar{t}}\left\{A_1(x(t),t)+B_1(x(t),t)ln\frac{1}{x^{(1)}(t)}\right\}\frac{d t}{t}\right]
\ee
and
\be
\label{eqn:ch6delv2tbar}
\Delta v^{(2)}(\bar{x},\bar{t})=\Delta v^{(2)}({\tau}'_2)\,\exp\left[\int_{t_0}^{\bar{t}}\left\{A_2(x(t),t)+B_2(x(t),t)ln\frac{1}{x^{(2)}(t)}\right\}\frac{d t}{t}\right]\, .
\ee
Or changing the variables from ($ \bar{x},\bar{t}$) to ($x,t$) 
\be
\label{eqn:ch6delv1t}
\Delta v^{(1)}(x,t)=\Delta v^{(1)}({\tau}'_1)\, \exp\left[\int_{t_0}^{t}\left\{A_1(x(t'),t')+B_1(x(t'),t)ln\frac{1}{x^{(1)}(t')}\right\}\frac{d t'}{t'}\right]
\ee
and
\be
\label{eqn:ch6delv2t}
\Delta v^{(2)}(x,t)=\Delta v^{(2)}({\tau}'_2)\, \exp\left[\int_{t_0}^{t}\left\{A_2(x(t'),t')+B_2(x(t'),t')ln\frac{1}{x^{(2)}(t')}\right\}\frac{d t'}{t'}\right]\, .
\ee
These two equations (Eq.(\ref{eqn:ch6delv1t}) and Eq.(\ref{eqn:ch6delv2t})) give the two unknown functions $\Delta v^{(1)}(x,t) $ and $\Delta v^{(2)}(x,t) $ which are related to the polarized singlet $\Delta \Sigma (x,t)$ and polarized gluon $ \Delta g (x,t) $ by Eq.(\ref{eqn:ch6vdef}). Solving Eq.(\ref{eqn:ch6vdef}) for $\vec {\Delta v}(x,t)$ we get
\be
\label{eqn:ch6delvrelation1}
\Delta v^{(1)}(x,t)=-\frac{(-b_{11}^\prime+b_{22}^\prime-s)}{2\,s}\Delta g(x,t)-\frac{b_{21}^\prime}{s}\Delta \Sigma (x,t)
\ee
and
\be
\label{eqn:ch6delvrelation2}
\Delta v^{(2)}(x,t)=\frac{(-b_{11}^\prime+b_{22}^\prime+s)}{2\,s}\Delta g(x,t)+\frac{b_{21}^\prime}{s}\Delta \Sigma (x,t)\, ,
\ee
where $ s $ is given by Eq.(\ref{eqn:ch6s}). From Eq.(\ref{eqn:ch6delvrelation1}) and Eq.(\ref{eqn:ch6delvrelation2}) we get
\be
\Delta \Sigma (x,t)=-\frac{(-b_{11}^\prime+b_{22}^\prime+s)}{2 b_{21}^\prime}\Delta v^{(1)}(x,t)-\frac{(-b_{11}^\prime+b_{22}^\prime-s)}{2 b_{21}^\prime}\Delta v^{(2)}(x,t)
\ee
and
\be
\label{eqn:ch6delg}
\Delta g(x,t)=\Delta v^{(1)}(x,t)+\Delta v^{(2)}(x,t)\,.
\ee
 In the special case when $x \rightarrow 0 $,
\be
\label{eqn:ch6delsigmax0}
\Delta \Sigma (x,t)=\frac{1}{4}(-8-\sqrt{55})\Delta v^{(1)}(x,t)+\frac{1}{4}(-8+\sqrt{55})\Delta v^{(2)}(x,t)\, 
\ee
and
\be
\label{eqn:ch6delgx0}
\Delta g(x,t)=\Delta v^{(1)}(x,t)+\Delta v^{(2)}(x,t)\,.
\ee
That is, at very low $x$ approximation
\bea
\label{eqn:ch6delsigmaxlow}
\Delta \Sigma (x,t)=&&\frac{1}{4}(-8-\sqrt{55})\Delta v^{(1)}({\tau}'_1)\nonumber \\
&&\times \exp\left[\int_{t_0}^{t}\left\{A_1(x(t'),t')+B_1(x(t'),t)ln\frac{1}{x^{(1)}(t')}\right\}\frac{d t'}{t'}\right]\nonumber \\
&&+\frac{1}{4}(-8+\sqrt{55})\Delta v^{(2)}({\tau}'_2)\nonumber \\
&&\times \exp\left[\int_{t_0}^{t}\left\{A_2(x(t'),t')+B_2(x(t'),t')ln\frac{1}{x^{(2)}(t')}\right\}\frac{d t'}{t'}\right]\nonumber \\
\eea
and
\bea
\label{eqn:ch6delgxlow}
\Delta g (x,t)=\Delta v^{(1)}({\tau}'_1)\exp\left[\int_{t_0}^{t}\left\{A_1(x(t'),t')+B_1(x(t'),t)ln\frac{1}{x^{(1)}(t')}\right\}\frac{d t'}{t'}\right]\nonumber \\
+\Delta v^{(2)}({\tau}'_2)\exp\left[\int_{t_0}^{t}\left\{A_2(x(t'),t')+B_2(x(t'),t')ln\frac{1}{x^{(2)}(t')}\right\}\frac{d t'}{t'}\right].
\eea
In Eq.(\ref{eqn:ch6delsigmaxlow})  and Eq.(\ref{eqn:ch6delgxlow}), $\Delta v^{(1)}({\tau}'_1) $ and $\Delta v^{(2)}({\tau}'_2) $ are the initial conditions on the initial curve $t=t_0$. Since  $\Delta v^{(1)} $ and $\Delta v^{(2)}$ are some combinations of $\displaystyle{\Delta \Sigma} $ and $\displaystyle{\Delta g} $ as given by Eq.(\ref{eqn:ch6delvrelation1}) and Eq.(\ref{eqn:ch6delvrelation2}), therefore, from the input distribution for $\displaystyle{\Delta \Sigma} $ and  $\displaystyle{\Delta g} $ at $\displaystyle{Q^2=Q_0^2}  $ ($ t=t_0$) , we can get $\Delta v^{(1)}({\tau}'_1) $ and $\Delta v^{(2)}({\tau}'_2) $ simply by substituting $x \rightarrow {\tau}'_1 $ and $x \rightarrow {\tau}'_2 $\cite{Farlow}. But in carrying out the integration we face one problem. Under the integral sign in Eq.(\ref{eqn:ch6delsigmaxlow}) and Eq.(\ref{eqn:ch6delgxlow}), we have the ratio $\displaystyle{\frac{v_1}{v_2}}$(see Eq.(\ref{eqn:ch6A1}-\ref{eqn:ch6B2})) which is yet unknown. This ignorance forbids analytical forms for the quantities defined in these equations. However, at the initial scale $t=t_0$ this ratio is known, because we know $\Delta \Sigma $ and $\Delta g$. We assume that the $x$ and $t$ dependent parts of $\Delta v^{(1)} $ and  $\Delta v^{(2 )}$ are factorisable and the $x $ dependent part does not deviate significantly from their values at $ t=t_0$ i.e.
\be
\label{eqn:ch6v2byv1}
\frac{\Delta v^{(2)}(x,t)}{\Delta v^{(1)}(x,t)}\approx \frac{\Delta v^{(2)}({\tau}'_2)}{\Delta v^{(1)}({\tau}'_1)}\Delta f(t)\, ,
\ee
where $\Delta f(t)$ is some unknown test function.
\subsection{Analytical solution when $ t $ is very near the boundary $t=t_0$}
It is possible to obtain some analytical forms for the function $\Delta \Sigma $ and $ \Delta g$ in a region very close to the boundary $t=t_0$. Expanding $\Delta f(t)$ in a Taylor series about $t=t_0$ and  retaining only the first term we get
\bea
\label{eqn:ch6ftexpand}
\Delta f(t)&=& \Delta f(t_0)+(t-t_0)\Delta f'(t)+......\\   \nonumber
& \approx & \Delta f(t_0) 
= \Delta f_0 \, .
\eea
With this approximation the integration of Eq.(\ref{eqn:ch6delv1t}) and Eq.(\ref{eqn:ch6delv2t}) give the following analytical forms
\be
\label{eqn:ch6delv1ana}
\hspace{0.4in}\Delta v^{(1)}(x,t)=\Delta v^{(1)}({\tau}'_1)\,\exp\left[B_{10}\,ln\frac{t}{t_0}\,ln\frac{1}{x}\right]\,\left(\frac{t}{t_0}\right)^{(A_{10}+B_{10})}{\tau'}_1^{(\frac{B_{10}{\tau}'_1}{\alpha^{(1)}})}\,x^{-(\frac{B_{10}{\tau}'_1}{\alpha^{(1)}})}
\ee
and
\be
\label{eqn:ch6delv2ana}
\hspace{0.4in} \Delta v^{(2)}(x,t)=\Delta v^{(2)}({\tau}'_2)\,\exp\left[B_{20}\,ln\frac{t}{t_0}\,ln\frac{1}{x}\right]\,\left(\frac{t}{t_0}\right)^{(A_{20}+B_{20})}{\tau'}_2^{(\frac{B_{20}{\tau}'_2}{\alpha^{(2)}})}\,x^{-(\frac{B_{20}{\tau}'_2}{\alpha^{(2)}})}\, ,
\ee
where
\bea
&A_{10}=&\frac{1}{2970}\left[(-2255+ 284\sqrt{55})+3(275- 52\sqrt{55})\frac{v_2({\tau}'_2)}{v_1({\tau}'_1)}\,\Delta f_0\right]\nonumber , \\
&B_{10}=&\frac{1}{2970}\left[(4400-404\sqrt{55})+4(-880+119\sqrt{55})\frac{\Delta v^{(2)}({\tau}'_2)}{\Delta v^{(1)}({\tau}'_1)}\,\Delta f_0\right] \nonumber , \\
&A_{20}=&\frac{1}{2970}\left[-(2255+ 284\sqrt{55})+3(275+52\sqrt{55})\frac{\Delta v^{(1)}({\tau}'_1)}{\Delta v^{(2)}({\tau}'_2)}\,\Delta f_0\right] \nonumber ,\\
&B_{20}=&\frac{1}{2970}\left[(4400+ 404\sqrt{55})-4(880+119\sqrt{55})\frac{\Delta v^{(1)}({\tau}'_1)}{\Delta v^{(2)}({\tau}'_2)}\,\Delta f_0\right],
\eea
\be
{\tau}'_1=x+\alpha^{(1)}\,ln(t_0/t)
\ee
and
\be
{\tau}'_2=x+\alpha^{(2)}\,ln(t_0/t)\, .
\ee
Using Eq.(\ref{eqn:ch6delv1ana}) and Eq.(\ref{eqn:ch6delv2ana}) in  Eq.(\ref{eqn:ch6delsigmaxlow}) and Eq.(\ref{eqn:ch6delgxlow}), we get the analytical solutions for the polarized singlet and the gluon distribution which are  valid when $x$ is very small and  $t$ is very close to the boundary  $t=t_0$ of perturbative evolution. Explicitly these are
\bea
\label{eqn:ch6singletana}
\Delta \Sigma (x,t)=\frac{1}{4}(-8-\sqrt{55})\Delta v^{(1)}({\tau}'_1)\exp\left[B_{10}\,ln\frac{t}{t_0}\,ln\frac{1}{x}\right]\,\nonumber \\
\times \left(\frac{t}{t_0}\right)^{(A_{10}+B_{10})}{\tau'}_1^{(\frac{B_{10}{\tau}'_1}{\alpha^{(1)}})}\,x^{-(\frac{B_{10}{\tau}'_1}{\alpha^{(1)}})} \nonumber \\
+\frac{1}{4}(-8+\sqrt{55})\Delta v^{(2)}({\tau}'_2)\exp\left[B_{20}\,ln\frac{t}{t_0}\,ln\frac{1}{x}\right]\,\nonumber \\
\times \left(\frac{t}{t_0}\right)^{(A_{20}+B_{20})}{\tau'}_2^{(\frac{B_{20}{\tau}'_2}{\alpha^{(2)}})}\,x^{-(\frac{B_{20}{\tau}'_2}{\alpha^{(2)}})}\nonumber \\
\eea
and
\bea
\label{eqn:ch6gluonana}
\Delta g(x,t)=\Delta v^{(1)}({\tau}'_1)\,\exp\left[B_{10}\,ln\frac{t}{t_0}\,ln\frac{1}{x}\right]\,\left(\frac{t}{t_0}\right)^{(A_{10}+B_{10})}{\tau'}_1^{(\frac{B_{10}{\tau}'_1}{\alpha^{(1)}})}\,x^{-(\frac{B_{10}{\tau}'_1}{\alpha^{(1)}})}\nonumber \\
+\Delta v^{(2)}({\tau}'_2)\,\exp\left[B_{20}\,ln\frac{t}{t_0}\,ln\frac{1}{x}\right]\,\left(\frac{t}{t_0}\right)^{(A_{20}+B_{20})}{\tau'}_2^{(\frac{B_{20}{\tau}'_2}{\alpha^{(2)}})}\,x^{-(\frac{B_{20}{\tau}'_2}{\alpha^{(2)}})}\, .\nonumber \\
\eea

\section{Results and discussion}
\label{sec:results}
In recent years there have been considerable growth of data \cite{PLAnthonyPLB493,SMCPRD58,SMCPRD70,HermesPLB404,HermesPLB442,HermesPRL84,CompassSchill} on inclusive polarized deep inelastic scattering of leptons off nucleons. These data on the polarized asymmetry provide indirect way of extracting the polarized parton distribution functions that are essential to understand the spin decomposition of the nucleon. The distributions are determined from a global QCD analysis of polarized experimental data using the DGLAP evolution equations. Several such distributions are available in the literature\cite{AAC00,LSSEPJC23,GRSV01,BlumleinBottcher01,Gehrmanref9,Ghoshetal}. We compare our results derived in the previous section with some of these exact solutions\cite{AAC00,LSSEPJC23,GRSV01} of DGLAP equations that are available in the hep database as Fortran codes\cite{DurhamHEPdb}.

We first compare  the polarized gluon distribution $\Delta G$ ($= x \Delta g $) given by Eq.(\ref{eqn:ch6gluonana}) with the exact distribution  AAC00\cite{AAC00} at $Q^2= 5 GeV^2$. The analytical expression  Eq.(\ref{eqn:ch6gluonana}), expected to be valid at very low $x$ and low $t$ has a free parameter $\Delta f_0$ introduced through  the Eq.(\ref{eqn:ch6ftexpand}). Since $\Delta f_0$ essentially represents the ratio of some combination of singlet to the gluon polarized distribution at the initial scale, we assume that its value cannot be zero or negative. Starting from a low positive value, we vary this parameter  till we get a reasonable agreement with the exact distributions. In  Fig.1(a) we show the graphs only for few representative values of $\Delta f_0$ along with the exact AAC00LO\cite{AAC00} distribution. To evolve our polarized gluon we also use the AACLO input  at the initial scale $Q_o^2=1GeV^2$ as given in the ref.\cite{AAC00}. From the figure(Fig.1(a)) we see that, while at $x<0.01$, there is no appreciable difference between  our distributions for different values of the parameter $\Delta f_0$, above $x>0.01$, the difference is significant. We can see that for $\Delta f_0 \approx 2.4$, our prediction conforms reasonably well with the exact distribution. However, if we change the input distributions, then the value of the parameter $\Delta f_0$ for which we get a reasonable agreement with the exact distribution also changes. In Fig.1(b-c) we compare our result with two other exact distributions: LSS01\cite{LSSEPJC23} and GRSV01\cite{GRSV01} at the same scale $Q^2=5 GeV^2$. For evolution of $\Delta g$ we use the inputs also from the above two references. From the figure (Fig.1(b)(c)) we see that agreement between our prediction and the exact solution  is good when $\Delta f_0 \simeq 2.2$ in the case of LSS input distribution and  $\Delta f_0 \simeq 2.0$ in the case of GRSV input  distribution. Thus we find that the value of $\Delta f_0$ is sensitive to the input distributions and different input distributions give different $\Delta f_0$ for a reasonable agreement with the exact ones. We use these  values of $\Delta f_0$ for the later part of our discussion to make the analysis self consistent.

 Now let us come to the gluon asymmetry $\frac{\Delta G}{G}$. Presently, the experimental data available on the gluon asymmetry are limited and there are only three data sets from SMC\cite{SMCPRD58}, HERMES\cite{HermesPLB404,HermesPLB442,HermesPRL84} and  COMPASS\cite{CompassSchill,CompassBarnet}  obtained from the asymmetry measurement of high-$p_T$ hadron production. 
\begin{sidewaysfigure}
\begin{center}
\subfigure[]{\includegraphics[width=3.5in]{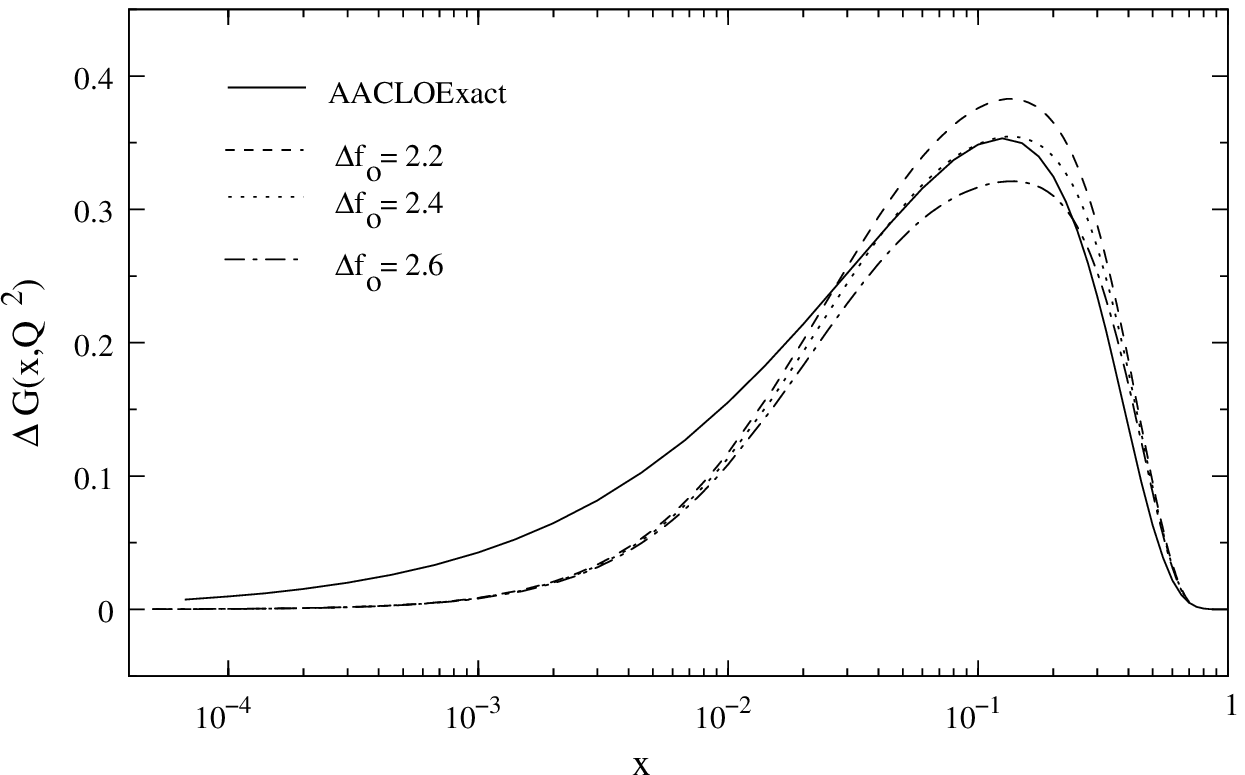}}
\subfigure[]{\includegraphics[width=3.5in]{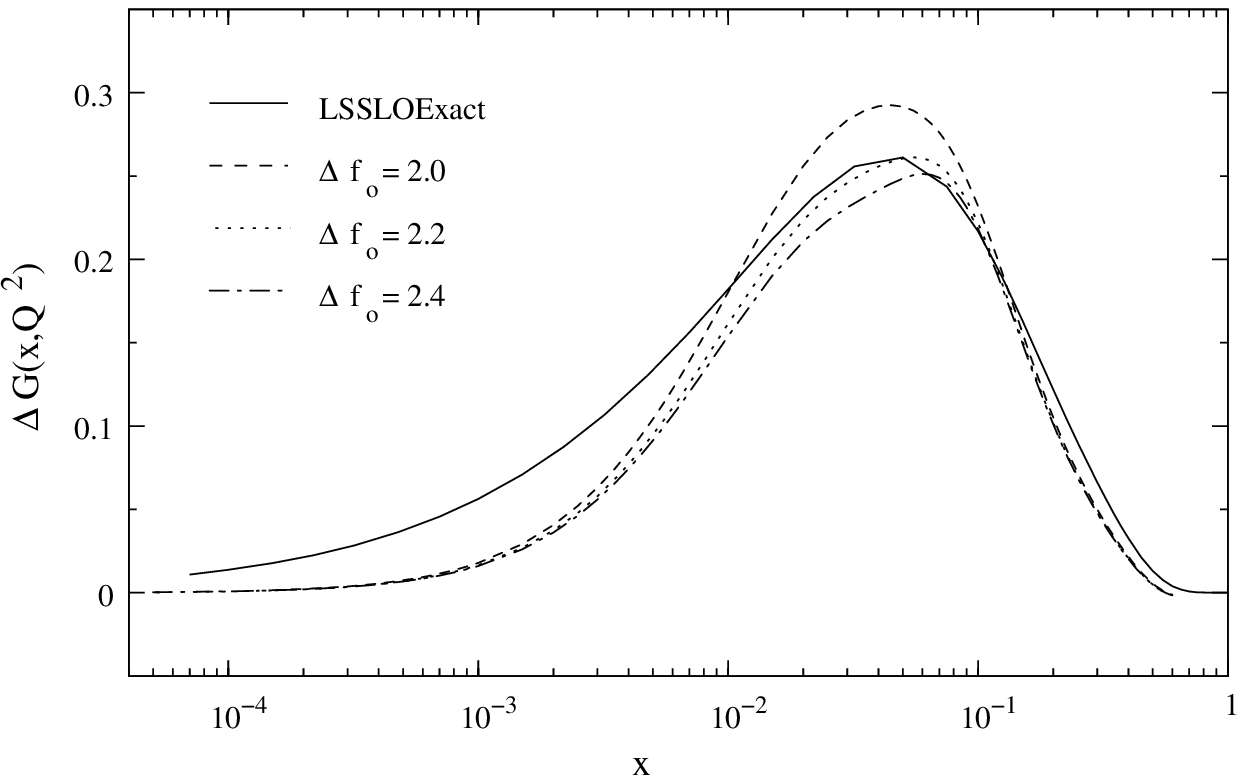}}\\
\subfigure[]{\includegraphics[width=3.5in]{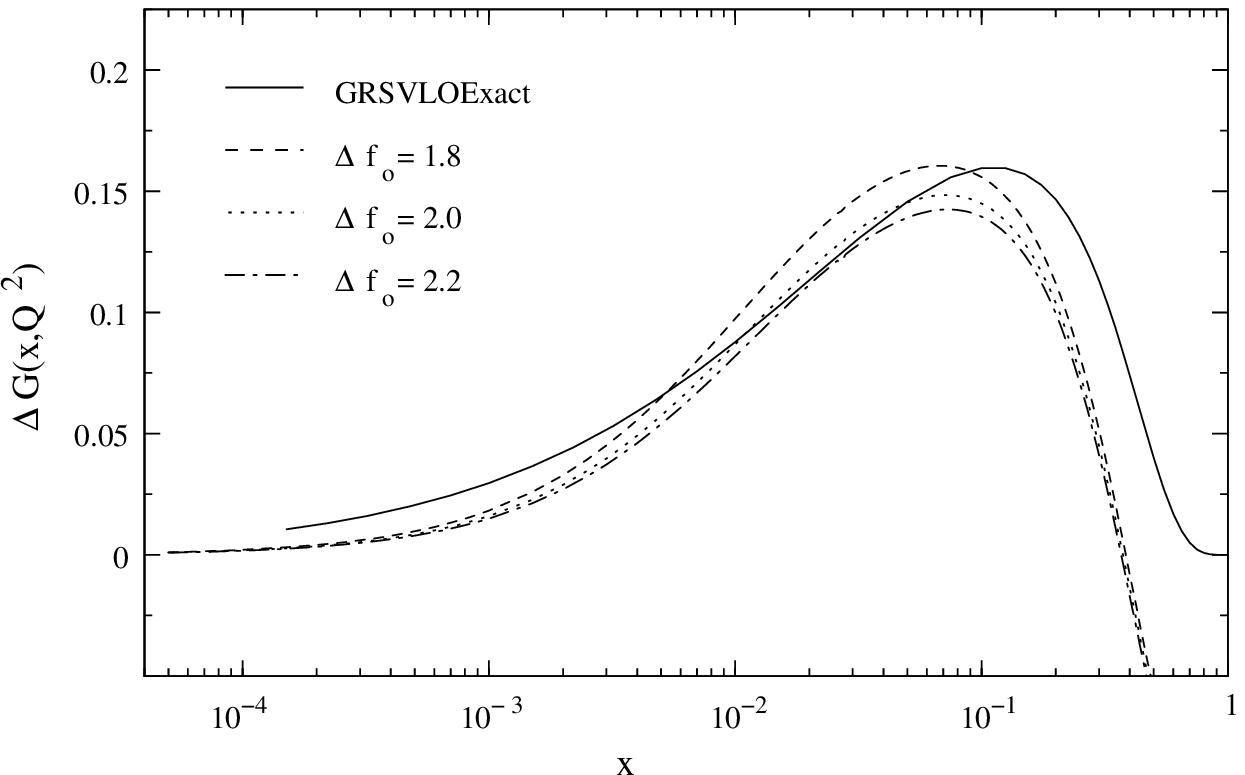}}
\end{center}
\caption[Polarized gluon given by the analytic expression Eq.(\ref{eqn:ch6gluonana}) compared with the exact  (a)AAC00LO, (b)LSS02LO and  (c) GRSV01LO distributions]{Polarized gluon given by the analytic expression Eq.(\ref{eqn:ch6gluonana}) compared with the exact  (a)AAC00LO\cite{AAC00}, (b)LSSLO\cite{LSSEPJC23} and  (c) GRSVLO\cite{GRSV01} distributions at $Q^2=5 GeV^2 $ for different values of the free parameter $\Delta f_0 $.}
\label{fig:ch6fig1}
\end{sidewaysfigure}
These are shown in table 1. The bracket $<>$ in the table  means that the value is obtained at an average fraction of the nucleon momentum carried by the struck quark. The measurement is carried out at a momentum scale of the order of $Q^2=2\sim 5 GeV^2$.
\begin{table}
\label{table:ch6tab1}
\begin{center}
\caption[Gluon asymmetry $\frac{\Delta G}{G}$ data from COMPASS, SMC and HERMES ]{Recent Gluon asymmetry $\frac{\Delta G}{G}$ data from COMPASS, SMC and HERMES}
\vspace{0.1in}
\begin{tabular}{lll}
\hline
COMPASS & $\frac{\Delta G}{G}= 0.06\pm 0.31(st)\pm 0.06(syst)$ & $ < x>=0.13$\\
SMC & $\frac{\Delta G}{G}= -0.20\pm 0.28(st)\pm 0.10(syst)$ & $ <x>=0.07$ \\
HERMES & $\frac{\Delta G}{G}= 0.41\pm 0.018(st)\pm 0.03(syst)$ & $ 0.06 < x < 0.28 $ \\\hline
\end{tabular}
\end{center}
\end{table}
We calculate the quantity $\frac{\Delta G}{G}$, with $\Delta G$ given by Eq.(\ref{eqn:ch6gluonana}) and the unpolarized gluon  distribution taken from the solution of the LO DGLAP gluon evolution equation  as obtained in \cite{PKSPram58}. Unpolarized gluon distribution is evolved with MRSTLO\cite{MRST2001} input distribution at $Q^2=1 GeV^2$. We show our calculated values for $\frac{\Delta G}{G}$ at $Q^2 = 5 GeV^2$  along with the available data  in Fig.2. The vertical error bar in the data represents the statistical error and the horizontal error bar, the standard deviation of the $x_g$ distribution at which the measurement is carried out. From the figure we see that our predicted $\frac{\Delta G}{G}$ reproduces the general trend of the values of $\frac{\Delta G}{G}$ indicated by the data. We also see that the gluon asymmetry $\frac{\Delta G}{G}$ invariably remains positive and conforms to the positivity constraint $\frac{\Delta G}{G}\leq 1$. However, this is true only at low momentum scale $Q^2\leq 5 GeV^2$. As the momentum scale is increased, the condition $\frac{\Delta G}{G} < 1$ breaks down at certain values of $x$ which we discuss in the next paragraph. 

At this point, few remarks about the positivity constraints are in order. The positivity condition has the origin in the probabilistic interpretation of the parton densities. According to it, the magnitude of a polarized cross-section should be smaller than the corresponding unpolarized one, \textit{i.e.} $|\Delta\sigma | \leq \sigma$. In LO, probabilistic interpretation can be applied to parton distributions and hence this condition (i.e. $|\Delta\sigma | \leq \sigma$) implies that the parton densities should satisfy the relation $$ \left|\Delta f\left(x,Q^2\right)\right| \leq f\left(x,Q^2\right).$$
\begin{figure}
\begin{center}
\includegraphics[width=5.0in]{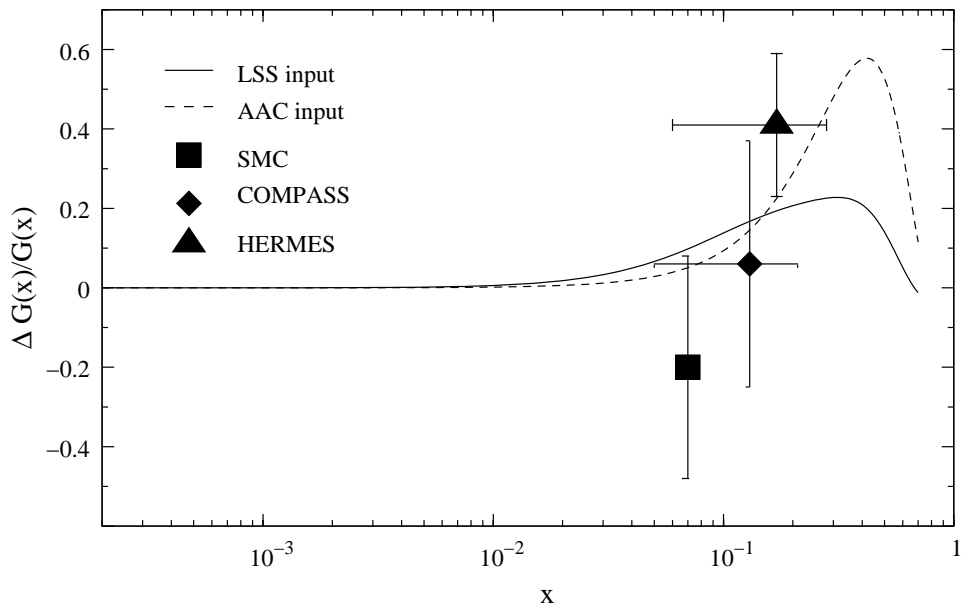}
\end{center}
\caption[Gluon asymmetry with two different inputs compared with data from SMC ,COMPASS and HERMES ]{ Gluon asymmetry with two different inputs compared with data from SMC\cite{SMCPRD70},COMPASS\cite{CompassSchill} and HERMES\cite{HermesPRL84}.Values of $\Delta f_0 $ are chosen from figure \ref{fig:ch6fig1} as discussed in the text.}
\label{fig:ch6fig2}
\end{figure}
In determining the  polarized distributions by different groups, the positivity condition is imposed along with other rules such as quark counting rule\cite{JFGunion,SJBrodosky,DRJackson} to find the parameters of the initial distributions. In our study of the polarized DGLAP equations, since we are not making any fit except varying the parameter $\Delta f_0$, we see for what kinematic regions, our solutions conform to the positivity condition  for a particular input distribution.
Taking the LSSLO\cite{LSSEPJC23} input distribution, we calculate $\frac{\Delta G}{G}$ for a large number of fixed $Q^2$ and see  for what value of $x$, if there is any, the condition $\frac{\Delta G}{G}<1$ breaks down. The value of the free parameter $\Delta f_0$ is chosen to be $2.2$ (another graph for $\Delta f_0 =2.0 $ also shown) as obtained from the previous analysis. In Fig.3 we show the $x-Q^2$ range within which the condition $\frac{\Delta G}{G}<1$ holds good. We see that as $Q^2$ is increased, the condition  $\frac{\Delta G}{G}<1$ breaks down at progressively lower values of $x$. On the other hand, for low momentum scale $ Q^2<5 GeV^2$ it is always less than one.
\begin{figure}
\begin{center}
\includegraphics[width=5.0in]{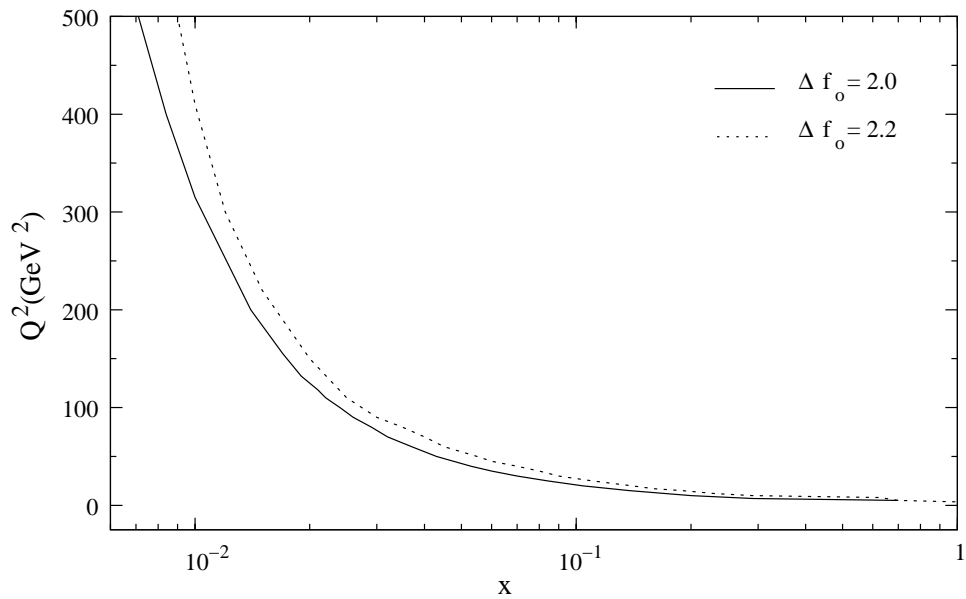}
\end{center}
\caption[$x-Q^2$ range where the positivity condition $\frac{\Delta G}{G} < 1$ holds good ]{$x-Q^2$ range where the positivity condition $\frac{\Delta G}{G} < 1$ holds good. In the region below the curve $\frac{\Delta G}{G} < 1$.The graphs are drawn with LSS\cite{LSSEPJC23} input  in Eq.(\ref{eqn:ch6gluonana}) for two different values of $\Delta f_0$. Unpolarized gluon $G(x,t)$ is taken from \cite{PKSPram58}.}
\label{fig:ch6fig3}
\end{figure}

Next we consider the flavour singlet polarized quark distribution $\Delta \Sigma\, (x,t)$ given by  Eq.(\ref{eqn:ch6singletana}) and compare it with exact  AACLO\cite{AAC00} and LSSLO\cite{LSSEPJC23} distributions. Taking the inputs also from AACLO\cite{AAC00} and LSSLO\cite{LSSEPJC23} respectively, we evolve  $\Delta \Sigma(x,Q^2)$ using Eq.(\ref{eqn:ch6singletana})  at two different scales $Q^2=2 GeV^2$ and $Q^2=5 GeV^2$ and plot these as  function of $x$ along with the exact solutions in Fig.4. The value of $\Delta f_0 $ is taken to be $2.2 $ for AAC input and $2.4 $ for LSS input respectively as determined earlier. From the graphs we see that the qualitative features of the LO exact distribution is well reproduced by the analytical form of $\Delta\Sigma(x,Q^2)$ given by Eq.(\ref{eqn:ch6singletana}).

The value of $\Delta\Sigma(x,Q^2)$ and $\Delta g (x,Q^2)$  also give us information about the  contribution of the quark and the gluon to the total spin of the proton. The contribution of the singlet quark $\Delta \Sigma \left(x,Q^2\right)$ and the gluon distributions $\Delta g(x,Q^2)$ to the total spin of the proton are given by their first moments:
$$\Delta\Sigma(Q^2) = \int_0^1\Delta\Sigma(x,Q^2)dx$$
$$\Delta g\left(Q^2\right) = \int_0^1\Delta g\left(x,Q^2\right)dx .$$
 \begin{sidewaysfigure}
\begin{center}
\subfigure[]{\includegraphics[width=3.0in]{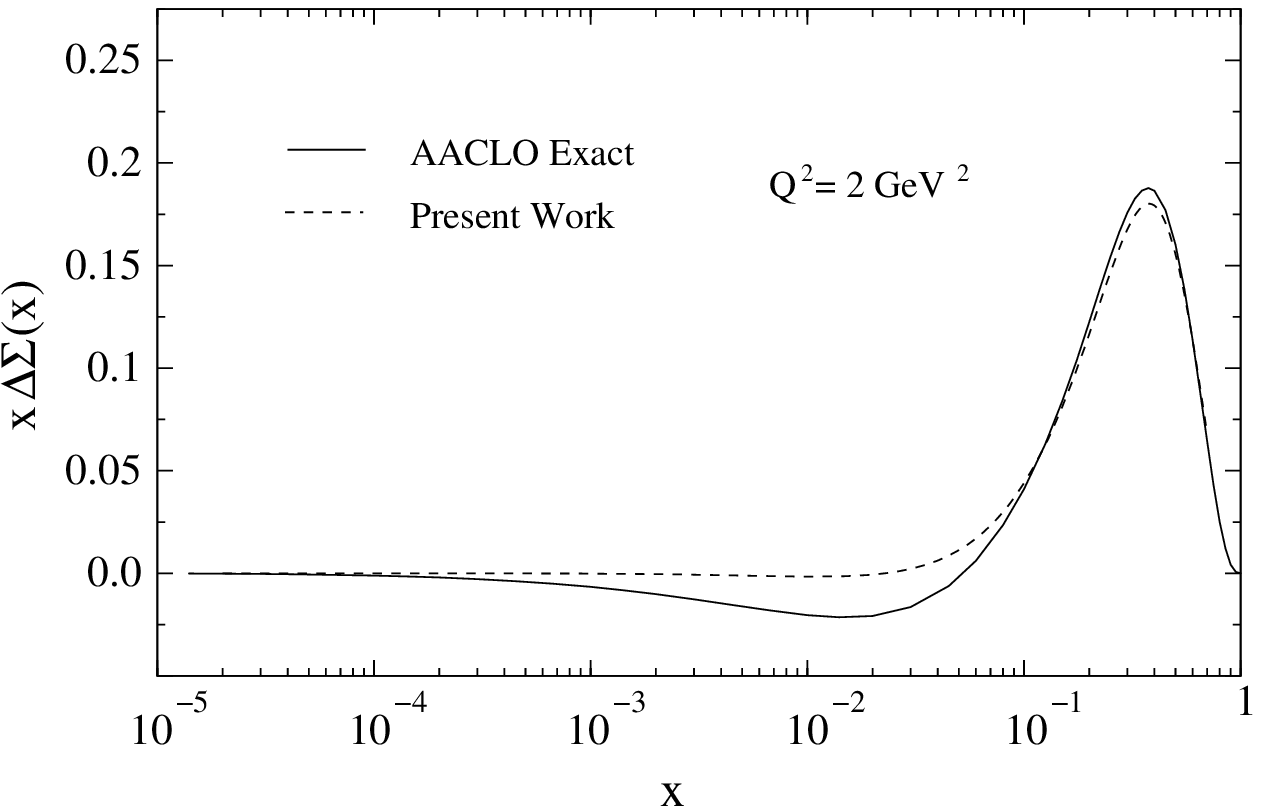}}
\subfigure[]{\includegraphics[width=3.0in]{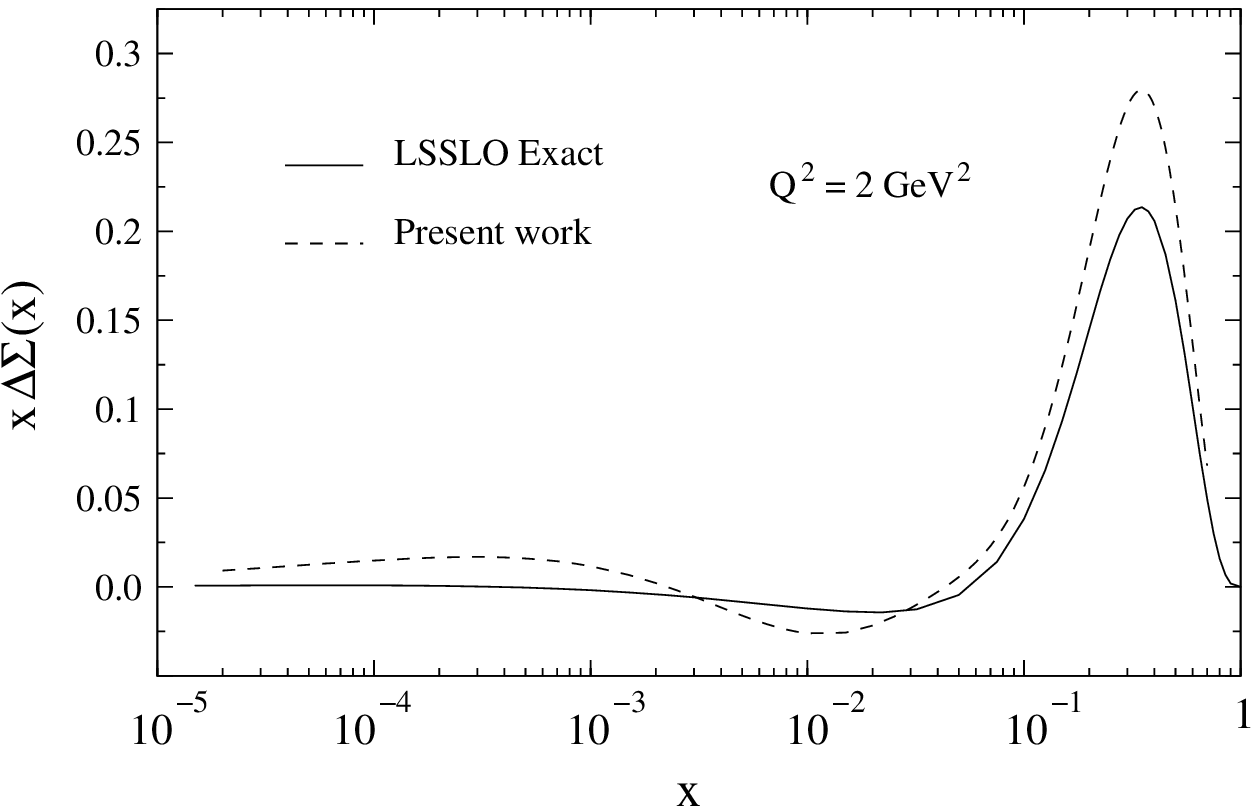}}\\
\subfigure[]{\includegraphics[width=3.0in]{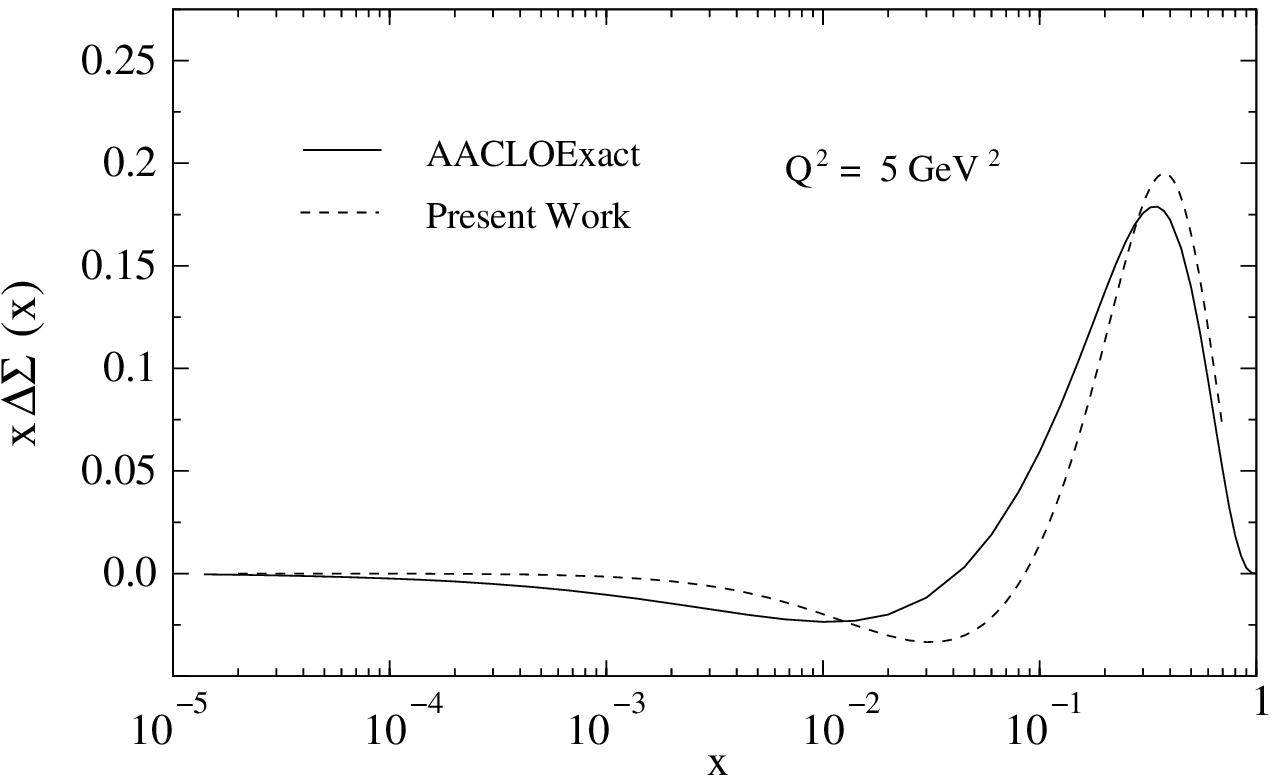}}
\subfigure[]{\includegraphics[width=3.0in]{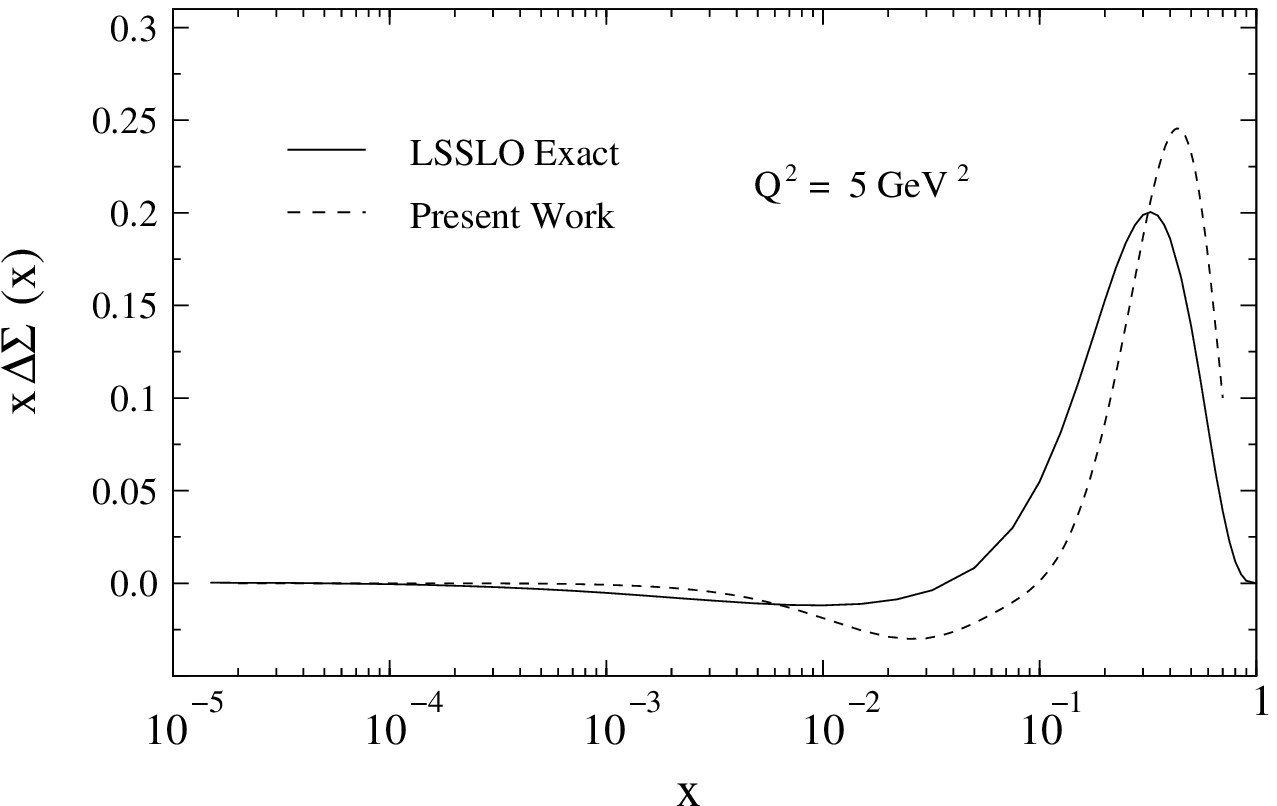}}
\end{center}
\vspace{-0.2in}
\caption[Polarized singlet distribution given by Eq.(\ref{eqn:ch6singletana}) as a function of $x$ compared with the exact AAC00LO and LSS02LO distributions at two fixed $Q^2=2 GeV^2$ and $Q^2= 5 GeV^2 $ ]{Polarized singlet distribution given by Eq.(\ref{eqn:ch6singletana}) as a function of $x$ compared with the exact AAC00LO\cite{AAC00} and LSSLO\cite{LSSEPJC23} distributions at two fixed $Q^2=2 GeV^2$ and $Q^2= 5 GeV^2 $. Values of $\Delta f_0 $ are chosen as discussed in the text.}
\label{fig:ch6fig4}
\end{sidewaysfigure}

To calculate the first moment we have to integrate over the entire range of $x$ from $0$ to $1$. But in our derivation, we have an upper limit of $x$ up to which we can calculate our functions $\Delta\Sigma(x,Q^2)$ and $\Delta g(x,Q^2)$. This limit is set by the functions ${\tau}'_1$ and ${\tau}'_2$ above which they exceed the physical limit of unity. This upper limit is dependent on $Q^2$. For $Q^2 = 5 GeV^2$, this upper limit is $x_{max}\approx 0.6$. We calculate the first moments for our functions $\Delta\Sigma(x,Q^2)$ and $\Delta g(x,Q^2)$ at $Q^2 = 5 GeV^2$ and show them in table 2  along with the results obtained by GRSV\cite{GRSV01} and AAC\cite{AAC00} LO parametrization. In table we have quoted only the LO result. The difference in their values for the spin content of the quark mainly originate from the small $x$ behaviour of the anti-quark distributions and the fact that accurate experimental data are still not available at small $x$. The usual range of $\Delta\Sigma$ obtained by different parametrization groups lies between $0.1\sim 0.3$ and we see that our calculated $\Delta\Sigma$ is within this range. For the gluon polarization we have obtained a low value compared to the two parametrizations, presumably due to the cut-off imposed on $x_{max}$ in evaluating the integrals. 

\begin{table}
\label{table:ch6tab2}
\begin{center}
\caption[Quark and Gluon helicity distribution ]{Quark and Gluon helicity distribution}
\vspace{0.1in}
\begin{tabular}{lllll}
\hline
 & $Q^2(GeV^2)$ & AAC & GRSV & ours\\
\hline
$\Delta\Sigma $& 5 & 0.18 & 0.259(standard) & 0.201\\
& & & 0.248(valence) & \\
\hline
$\Delta$g & 5 & 1.314 & 0.684 (standard) & 0.31\\
& & & 0.963 &\\
\hline
\end{tabular}
\end{center}
\end{table}

For completeness, we also note that a parallel analysis is possible if one assumes an \textit{ad-hoc} analytical form of the unknown function $\Delta f(t)$ in Eq.(\ref{eqn:ch6v2byv1}) instead of the parameter $\Delta f_0$ (Eq.(\ref{eqn:ch6ftexpand})) adjusted to the exact result. We refrain from doing such an analysis since $Q^2$ explored in the PDIS is limited near the boundary $t\approx t_o$ and effectively $\Delta f(t) \simeq \Delta f(t_0)=\Delta f_0$.

\end{document}